\def\L{{\cal L}}
\documentstyle[12pt]{article}
\begin{document}
\begin{flushright}
DO-TH/92-24\\
December 1992
\end{flushright}
\vspace{2.cm}
\centerline{\LARGE \bf Multi-Higgs Doublet Model:}
\vspace{0.5cm}
\centerline{\LARGE \bf Can It Account for $l^+l^-\gamma\gamma$ Events ?}
\vspace{2.cm}
\begin{center}
{{\large G.~Cveti\v c, M.~Nowakowski and Yue-Liang Wu} \\
Inst.~f\"ur Physik, Universit\"at Dortmund, 4600 Dortmund 50, Germany}
\end{center}
\vspace{2.cm}
\centerline{\bf ABSTRACT}

Motivated by recent experiments at LEP ($Z \rightarrow l^+l^-\gamma\gamma$,
with $M_{\gamma\gamma} \simeq 59 GeV$), we examine the possibility that
such events are explained by a Bjorken process with a resonant Higgs
($M_{h^o} \simeq 59 GeV$) decaying to $\gamma\gamma$. Our model is an
N-doublet-Higgs model without CP-violation. Under the simplifying
assumption that $(N-k)$ doublets ($k \geq 1$) decouple from the fermion
sector, we could explain the observed events if: i) $(N-1) \geq 2$ and
ii) the masses of the corresponding $(N-1)$ neutral scalars are either
completely or almost degenerate, i.~e.~ the decay $Z \rightarrow l^+l^-
\gamma\gamma$ proceeds now via several overlapping resonances.

\newpage
\vspace{1cm}

Recently, the L3 Collaboration has found four events $Z \rightarrow
l^+l^-\gamma\gamma$ with the invariant mass of the photon pairs
$M_{\gamma\gamma} \simeq 59 GeV$ ~\cite{L3}.
The total number of the corresponding analyzed Z-decays is
$N_{tot}=1.6 \times 10^6$ ~\cite{Pohl} (up until mid-November 1992).

In this letter, we investigate the possibility that this process is
realized via the Bjorken process $Z \rightarrow Z^{\ast}h^o$, where
$Z^{\ast} \rightarrow l^+l^-$ and $h^o \rightarrow \gamma\gamma$
($M_{h^o} \simeq 59 GeV$), within the simplest possible extensions
of the minimal standard model (SM) - those with $N$ scalar isodoublets
($N \geq 2$). The observed decays can't be explained within
the minimal SM ($N = 1$), among other things because the branching
ratios $\Gamma (h^o \rightarrow \gamma\gamma )/ \Gamma (h^o
\rightarrow \mbox{all} ) \ll 1$ (e.~g.~, the decays $h^o \rightarrow
\bar b b$ would be much more frequent than $h^o \rightarrow \gamma
\gamma$ ).

There are, in general, three problems when one tries to explain
these high photon mass events by a Bjorken process. Each of these
problems is connected to one of the vertices in the decay diagram.
\begin{enumerate}
\item The first problem is the deficit of the $Z \rightarrow \nu
\bar \nu \gamma\gamma$ and $Z \rightarrow q \bar q \gamma\gamma$
events, relative to $Z \rightarrow l^+l^- \gamma \gamma$ events.
Since the experimental results might be too tentative to draw
definite conclusions on this point, we will not discuss it here.
\item The branching ratio $B(h^o \rightarrow \gamma\gamma)$ should
be of the order one. This is excluded in the minimal SM, but could
be arranged in a multi-doublet-Higgs model (see below).
\item The third problem concerns the on-shell production rate for
$Z \rightarrow Z^{\ast}h^o \rightarrow l^+l^-h^o$.
In SM, one predicts $1.3$ such events if
$M_{h^o} \simeq 59 GeV$ and the total number of $Z$~-decays is
$1.6 \times 10^6$. This would make it difficult to reconcile the
observed four events with the theory, even when $B(h^o \rightarrow
\gamma\gamma) \simeq 1$.
\end{enumerate}

Below, we will concentrate primarily on the issue 3., taking the
optimal assumption that $B(h^o_j \rightarrow \gamma\gamma)$ is roughly
one for all neutral ($CP=+1$) Higgses contributing to the process.

If we allow for $N \geq 2$, we can in principle find a scenario in
which only one neutral ($CP = +1$) scalar (for example, the Nth one)
couples to fermions, while all the other
neutral scalars ($\{ h^o_j \}, j = 1, \ldots , N-1$) do not. The
decoupling of $(N-1)$ neutral scalar Higgses from the fermion sector
could ensure that $\Gamma(h^o_j \rightarrow \gamma\gamma)$ is one
of the dominant decay channels. In order to have
$B(h^o_j \rightarrow \gamma\gamma) \simeq 1$, one has to tune the
parameters of the theory. We will find that even under these optimal
conditions, the experimental result can be explained by the Bjorken-type
process only if the $N$~-doublet model displays some ``exotic'' features
like degeneracy of masses.

Let's denote the $N$ doublets $\Phi_j$ ($j=1, \ldots, N$) as:
\begin{eqnarray}
\Phi_j & = & \frac{e^{i \delta_j}}{\sqrt{2}}
      \left( \begin{array}{c}
     \phi^{(1)}_j + i \phi^{(2)}_j \\
     v_j + H^o_j + i A^o_j
     \end{array} \right) \,
   \nonumber\\
\langle \Phi_j \rangle_o & = & \frac{e^{i \delta_j}}{\sqrt{2}}
     \left( \begin{array}{c}
     0 \\ v_j \end{array} \right) \ \cdot
\end{eqnarray}

{}From the gauge boson-scalar interaction part
\begin{equation}
\L_{gb-sc} = \sum_{j=1}^N \left( D_{\mu} \Phi_j \right)^{\dagger}
\left( D^{\mu} \Phi_j \right)
\end{equation}
we then obtain
\begin{equation}
\L^{ZZ-sc} = \frac{1}{4} (g^2+g'^2) \left[ 4\sum_{j=1}^N v_jH^o_j \right]
 Z^{\mu}Z_{\mu} \,
\end{equation}
where the $Z$-mass term yields the condition
\begin{equation}
\sum_{j=1}^Nv^2_j=v^2 \qquad (=(246 GeV)^2) \ \cdot
\end{equation}

Assuming that there is no CP-violation (or: that it is reasonably small),
i.~e.~ $\delta_j$~'s are zero, then the physical $CP=+1$ scalars $h^o_j$
are obtained by an orthonormal transformation $O$ of $H^o_j$~'s, determined
by the scalar potential $V( \{ \Phi_j \} )$ (there are no substantial
admixtures of $A^o_j$~'s which have $CP=-1$)
\begin{equation}
h^o_j  =  O_{ji}H^o_i \ \ ,
 \qquad H^o_j  =  O_{ij}h^o_i  \ \cdot
\end{equation}
We assume that $O$ is a diagonal block matrix made up of one
$(N-1) \times (N-1)$ and one $1 \times 1$ diagonal block, i.~e.~
that the $(CP=+1)$ neutral scalar component of the doublet $\Phi_N$
(-the latter couples to fermions) doesn't have quadratic mixing terms
with the corresponding components of the doublets
$\{ \Phi_j \}_{j=1, \ldots, N-1}$ in the Higgs potential
$V( \{ \Phi_j \}_{j=1, \ldots, N})$.

If we assume that there is just one $h^o_j$ that has
$M_{h^o_j} \simeq (59 \pm 1) GeV$ ($j$ is fixed, $1 \leq j \leq N-1$),
then we have the following relations for the process
$Z \rightarrow Z^{\ast}h^o_j \rightarrow l^+l^-\gamma\gamma$
\begin{eqnarray}
\frac{N_{events}}{N_{decays}} & = &
  \frac{\sum_{l=e,\mu} \Gamma \left( Z \rightarrow Z^{\ast}h^o_j
 \rightarrow l^+l^-\gamma\gamma \right) }
  {\Gamma \left( Z \rightarrow \mbox{all} \right) }
 \nonumber\\
 & = &  \left( \frac{g_{h^o_jZZ}}{g^{SM}_{h^oZZ}} \right)^2
 \left[ \frac{\sum_{l=e,\mu} \Gamma^{SM} \left( Z \rightarrow
 l^+l^-h^o \right) }{ \Gamma \left( Z \rightarrow \mbox{all} \right)}
 \right] \left[ \frac{\Gamma \left( h^o_j \rightarrow \gamma\gamma \right)}
 {\Gamma \left( h^o_j \rightarrow \mbox{all} \right) } \right]
 \nonumber\\
 & = & \kappa_j ( 0.84 \pm 0.13) \times 10^{-6} B(h^o_j \rightarrow
 \gamma\gamma) \leq \kappa_j (0.84 \pm 0.13) \times 10^{-6} \ \cdot
\end{eqnarray}
The superscript $SM$ denotes the corresponding quantities in the minimal
SM, with $M_{h^o}= (59 \pm 1) GeV$. Furthermore, we denoted
\begin{displaymath}
\kappa_j = \left( \frac{g_{h^o_jZZ}}{g^{SM}_{h^oZZ}} \right)^2 \ \cdot
\end{displaymath}
The values of the ratio
\begin{displaymath}
\frac{\sum_{l=e,\mu} \Gamma^{SM} \left( Z \rightarrow
 l^+l^-h^o \right) }{ \Gamma \left( Z \rightarrow \mbox{all} \right) }
= (0.84 \pm 0.13) \times 10^{-6}
\end{displaymath}
were obtained by using known formulas of the minimal SM (e.~g.~\cite{g}),
where the upper bound corresponds to $M_{h^o} = 58 GeV$ and the
lower bound to $M_{h^o}=60 GeV$.

For $N \geq 2$ case, there are several possible ways to increase
$\Gamma(h^o_j \rightarrow \gamma\gamma)$ ~\cite{kw}, by increasing the
contribution of the decay $h^o_j \rightarrow \gamma\gamma$ proceeding
via loops with charged Higgses. In such a case, we may expect
\begin{displaymath}
B \left( h^o_j \rightarrow \gamma\gamma \right) =
\frac{\Gamma \left( h^o_j \rightarrow \gamma\gamma \right) }
{\Gamma \left( h^o_j \rightarrow \mbox{all} \right) } \sim 1 \ ,
\end{displaymath}
because the decays $h^o_j \rightarrow \bar f f$ ($j \leq N-1$)
are not allowed at the
tree level, and because the decays $h^o_j \rightarrow W^{+\ast}
W^{-\ast}, Z^{\ast}Z^{\ast}$ contribute in general at most a few
percent to $\Gamma(h^o_j \rightarrow \mbox{all})$.
Namely, the latter couplings are of similar strength as those in the
minimal SM where $\Gamma(h^o \rightarrow W^{+\ast}W^{-\ast})$ and
$\Gamma(h^o \rightarrow Z^{\ast}Z^{\ast})$ have been calculated
{}~\cite{lhc}. The inequality in (6) could in such a case be approximated
by equality if the branching ratio $B(h^o_j \rightarrow \gamma\gamma)$
is close to one. However, the factor $\kappa_j$ in (6) is severely bounded
from above:
\begin{equation}
\kappa_j = \left( \frac{g_{h^o_jZZ}}{g^{SM}_{h^oZZ}} \right)^2
= \left( \sum_{i=1}^N O_{ji} \frac{v_i}{v} \right)^2 =
\left[ \frac{ \left( v^{rot} \right)_j}{v} \right]^2 \leq 1 \ ,
\end{equation}
because
\begin{equation}
\sum_{j=1}^Nv^2_j = \sum_{j=1}^N \left( v^{rot} \right)^2_j
 = v^2 \quad = (246 GeV)^2 \ \cdot
\end{equation}
Therefore, in such a case we obtain
\begin{displaymath}
\frac{\sum_{l=e,\mu}N \left( Z \rightarrow l^+l^-\gamma\gamma \right) }
{N \left( Z \rightarrow \mbox{all} \right) } \vert_{M_{\gamma\gamma}
\simeq 59 GeV} \leq \kappa_j (0.84) \times 10^{-6} < 0.84 \times 10^{-6}
\ \cdot
\end{displaymath}
This would predict for $N(Z \rightarrow \mbox{all}) = 1.6 \times 10^6
\mbox{events}$
\begin{equation}
\langle N \left( Z \rightarrow l^+l^-\gamma\gamma \right) \rangle
\vert_{M_{\gamma\gamma} \simeq 59 GeV} < 1.33 \ \cdot
\end{equation}
Since the L3 Collaboration has found four such events (among $1.6 \times
10^6$ Z-decays), such a case appears to be unlikely, although not
excluded.

However, there exists a scenario which would sufficiently increase the
upper bound on the r.~h.~s.~of relation (9), by increasing the ``effective''
$\kappa$ in relation (6) beyond $1$. Namely, let's assume that all (or some)
of those scalars $h^o_j$ which don't couple to fermions are degenerate
in masses. For simplicity, we take that all $h^o_j$ ($j=1, \ldots, N-1$)
are either completely or almost degenerate ($h^o_N$ is the only neutral
scalar with $CP=+1$ that couples to fermions at the tree level)
\begin{eqnarray}
M_{h^o_j} & \simeq & 59 GeV  \qquad (j=1, \ldots, N-1) \ ,
\nonumber\\
\mid \triangle M_{h^o_j} \mid & < & \Gamma_{h^o_j} \ ,
\end{eqnarray}
\footnote{The constraint on the mass difference is required for
the overlapping of the $(N-1)$ resonances.}
and that their decay amplitudes to two photons are approximately equal.
Hence, also the amplitudes
\begin{displaymath}
\frac{A \left( Z \rightarrow Z^{\ast}h^o_j \rightarrow l^+l^-\gamma\gamma
\right) }{g_{h^o_jZZ}} =
\frac{A_j}{g_{h^o_jZZ}} \qquad (j=1, \ldots, N-1)
\end{displaymath}
would be approximately equal:
\begin{equation}
\frac{A_1}{g_{h^o_1ZZ}} \simeq \frac{A_2}{g_{h^o_2ZZ}}
\simeq \cdots \simeq \frac{A_{N-1}}{g_{h^o_{N-1}ZZ}} \ \cdot
\end{equation}
In such a case, we would obtain
\begin{equation}
\Gamma \left( Z \rightarrow l^+l^-\gamma\gamma \right)
\vert_{M_{\gamma\gamma} \simeq 59GeV} \simeq
\left[ \frac{\Gamma \left( Z \rightarrow Z^{\ast}h^o_i \rightarrow
l^+l^-\gamma\gamma \right) }{ \left( g_{h^o_iZZ} \right)^2} \right]
\left( \sum_{j=1}^{N-1}g_{h^o_jZZ} \right)^2 \ \ ,
\end{equation}
where $i$ is any (fixed) index between $1$ and $N-1$.
Since the expression in the $[ \cdots ]$~-brackets is independent of
the $g_{h^o_iZZ}$ couplings, we effectively get in such a case in the
relation (6)
\begin{equation}
\kappa_j \longmapsto \kappa_{eff} = \left[ \sum_{j=1}^{N-1}
 \frac{g_{h^o_jZZ}}{g^{SM}_{h^oZZ}}  \right]^2
= \left[ \sum_{j=1}^{N-1} \frac{ \left( v^{rot} \right)_j}{v} \right]^2
\ \cdot
\end{equation}
Due to the ``Z-mass'' constraint (8), we get for a given $v_N$
($= O_{Ni}v_i = (v^{rot})_N$) the maximum of $\kappa_{eff}$ at
\begin{eqnarray}
\frac{ \left( v^{rot} \right)_1}{v} & = & \cdots  =
\frac{ \left(v^{rot} \right)_{N-1}}{v}
\nonumber\\
& = & \frac{1}{\sqrt{N-1}} \sqrt{1- \left( \frac{v_N}{v} \right)^2} \ ,
\end{eqnarray}
\begin{equation}
\kappa^{max}_{eff} = (N-1) \left[1 -
\left( \frac{v_N}{v} \right)^2 \right] \leq (N-1) \ \cdot
\end{equation}
Therefore, in such a scenario, we have instead of relation (6)
the following relation for the number of events
\begin{eqnarray}
\frac{ \sum_{l=e,\mu}N \left( Z \rightarrow l^+l^-\gamma\gamma \right) }
{N \left( Z \rightarrow \mbox{all} \right) } \vert_{M_{\gamma\gamma}
\simeq 59 GeV} & = & \kappa^{max}_{eff} \left[ (0.84) \times 10^{-6} \right]
B \left( h^o_i \rightarrow \gamma \gamma \right)
\nonumber\\
 & \leq & (N-1) \left[ 1 - \left( \frac{v_N}{v} \right)^2 \right]
 \left[ (0.84) \times 10^{-6} \right] \times 1
\nonumber\\
 & < & (N-1) (0.84) \times 10^{-6} \ \cdot
\end{eqnarray}
The number of expected events is thus enhanced by a factor of $(N-1)$,
due to the degeneracy of $(N-1)$ neutral scalar Higgses which makes the
summation of their amplitudes become coherent.

This scenario can be realized by the following N-Higgs-doublet model.
We introduce $N$ complex $SU(2)_L$~-doublet scalar fields
$\Phi_i$ ($i=1, \ldots, N$) with $Y=1$, where only $\Phi_N$ is the
standard Higgs doublet that couples to fermions (but with VEV $v_N <
v = 246 GeV$). We assume that the other $(N-1)$ doublets decouple
from fermions, by requiring, for example, a global $U(1)$ symmetry
$\Phi_i \rightarrow e^{i \alpha} \Phi_i$ ($i=1, \ldots, N-1$). A
Higgs potential which spontaneously breaks $SU(2)_L \times U(1)_Y$
down to $U(1)_{em}$ and has degenerate mass for the $N-1$ extra neutral
scalar Higgses can be written as
\begin{eqnarray}
\lefteqn{ V \left( \{ \Phi_i \} \right)  =  \lambda_N
\left( \Phi_N^{\dagger}\Phi_N - v_N^2 \right)^2 + }
\nonumber\\
 & & + \lambda_1 \sum_{i=1}^{N-1} \left( \Phi_i^{\dagger}
\Phi_i - v'^2 \right)^2 +
\nonumber\\
 & & + \lambda_2 \sum_{i<j}^{N-1} \left[ \left( \Phi_i^{\dagger} \Phi_j +
\Phi_j^{\dagger} \Phi_i - 2 v'^2 \right)^2 - 2 \left( \Phi_i^{\dagger}
\Phi_i - v'^2 \right) \left( \Phi_j^{\dagger} \Phi_j - v'^2 \right) \right]
\nonumber\\
 & & + \sum_{i<j}^{N-1} \lambda_{ij} \left( \Phi_i^{\dagger} \Phi_i
 \Phi_j^{\dagger} \Phi_j - \Phi_i^{\dagger} \Phi_j \Phi_j^{\dagger}
\Phi_i \right)
\nonumber\\
 & & - \sum_{i<j}^{N-1} \lambda^{'}_{ij} \left( \Phi_i^{\dagger} \Phi_j -
\Phi_j^{\dagger} \Phi_i \right)^2 \ ,
\end{eqnarray}
where all the $\lambda_i, \lambda_{ij}$ and $\lambda^{'}_{ij}$ are real
parameters, the potential is bounded from below, and the minimum of the
potential is manifestly at
\begin{equation}
\langle \Phi_N \rangle_o = { 0 \choose v_N } \, \qquad
\langle \Phi_i \rangle_o = { 0 \choose v'} \quad (i=1, \ldots, N-1) \ \cdot
\end{equation}
In this potential, the neutral ($CP=+1$) Higgs masses are given by
\begin{eqnarray}
M^2_{h^o_i} & = & 4 v'^2 [ \lambda_1 + (N-2) \lambda_2 ] \qquad
(i=1, \ldots, N-1)
\nonumber\\
M^2_{h^o_N} & = & 4 v^2_N \lambda_N \ \cdot
\end{eqnarray}
The charged Higgs masses are determined by $\lambda_{ij}$~'s, and the
``pseudoscalar'' Higgs ($CP=-1$) masses by $\lambda^{'}_{ij}$~'s.

A soft breaking of the degeneracy could occur by replacing $\lambda_1$,
$v'$ and $\lambda_2$ in (17) by slightly doublet-dependent values, and
by adding (small) terms to the potential (17):
\begin{equation}
\triangle V = \sum_{i<j}^{N-1} \lambda^{(3)}_{ij}  \left( \Phi_i^{\dagger}
\Phi_i -(v'_i)^{2} \right) \left( \Phi_j^{\dagger} \Phi_j - (v'_j)^{2} \right)
\ ,
\end{equation}
where
\begin{displaymath}
\mid \lambda^{(3)}_{ij} \mid \ll \mid \lambda_1 \mid , \
\mid \lambda_2 \mid \quad \mbox{and} \quad
 \mid v'_i-v' \mid \ll v' \ \cdot
\end{displaymath}
In order to have the enhancement mechanism leading to (16) also in such
a case, the resulting mass differences should satisfy $\mid \triangle
M_{h^o_j} \mid < \Gamma_{h^o_j}$, in order for the coherence conditions (11)
to survive.

Note that, according to the relation (16),
the described mechanism would predict for $N_{tot}=1.6 \times 10^6$
$Z$~-decays (and for $\sqrt{1- \left( \frac{v_N}{v} \right)^2} \simeq 1$)
the number of $(Z \rightarrow Z^{\ast}h^o \rightarrow l^+l^-\gamma\gamma)$
events (with $M_{\gamma\gamma} = M_{h^o} \simeq 59 GeV$)
as the following function of $(N-1)$ (-the number
of Higgs doublets which don't couple to fermions):
\begin{eqnarray}
\langle N_{events} \rangle  \simeq 1.35 \times (N-1)
\nonumber \\
\mbox{for} \quad N_{tot} = 1.6 \times 10^6 \quad \mbox{and} \quad
M_{\gamma\gamma} = 59 GeV \ \cdot
\end{eqnarray}

\begin{tabular}{rr}
\hline
$N-1 = 1$ & $\langle N_{events} \rangle \simeq 1.35$ \\
$    = 2$ &                            $\simeq 2.69$ \\
$    = 3$ &                            $\simeq 4.04$ \\
$    = 4$ &                            $\simeq 5.38$ \\
\hline
\end{tabular}

If we now relax our condition that the branching ratios
$B(h^o_j \rightarrow \gamma\gamma)$ for the $(N-1)$
degenerate Higgses are the same and of order one, then the required number
of Higgs doublets would increase. This would make an already exotic
model even more exotic.

There are several other experiments at LEP (DELPHI, ALEPH, OPAL) which will
possibly also yield the events $Z \rightarrow l^+l^-\gamma\gamma$ with
$M_{\gamma\gamma} \simeq 59 GeV$ (out of $1.6 \times 10^6$ Z-decays).
However, due to small number of these events, the possible
statistical fluctuations are and will probably remain
quite large. The four events of the L3 Collaboration ~\cite{L3} would
suggest, within our proposed scenario, that the number of the
degenerate neutral ($CP=+1$) scalars (with the mass $ \simeq 59 GeV$)
is two or more (i.~e.~ $(N-1) \geq 2$).

Other authors ~\cite{b} have discussed a similar scenario, with
$N-1 = 1$ (no degeneracy).

According to the model, the LEP-experiments should observe also
events with two photons and missing energy (due to $\bar \nu \nu$),
and events with $\bar q q \gamma\gamma$, at $M_{\gamma\gamma}
\simeq 59 GeV$.

The arguments of this paper remain basically unchanged when we assume
that $k$ scalars $h^o_j$($k \geq 1$; $j=N-k+1, \ldots, N$) couple to
fermions at the tree level, and that $(N-k-l)$ ($l \geq 0$) scalars
$h^o_j$ ($j=1, \ldots, N-k-l$) are degenerate at $M \simeq 59 GeV$.
In such a case, the factor $(N-1)$ in the relation (16) would be replaced
by $(N-k-l)$ and $\left[1 - \left( \frac{v_N}{v} \right)^2 \right]$ by
$\left[ 1 - \sum_{j=N-k-l+1}^N \frac{ \left( v^{rot} \right)_j^2}{v^2}
\right]$.
In such a case, the L3 results would suggest
$(N-k-l) \geq 2$. Such a more general scenario would allow for
additional Bjorken processes at $l$ other resonant energies
$M_{\gamma\gamma} = M_{h^o_j}$ ($j=N-k-l+1, ..., N-k$), with the
corresponding $\kappa_j = \left[ \frac{ \left( v^{rot} \right)_j}
{v} \right]^2 \ (<1)$ in the relation (6).
\\
\\[1cm]
{\bf Conclusions and Discussions}
If the four events $Z \rightarrow l^+l^-\gamma\gamma$ ($M_{\gamma\gamma}
\simeq 59 GeV$) detected by L3 Collaboration ~\cite{L3} are Bjorken-type
processes $Z \rightarrow Z^{\ast}h^o \rightarrow l^+l^-\gamma\gamma$ and
if the theory behind them is a (minimal) extension of SM with several ($N$)
Higgs doublets and no (or small) CP-violation in the Higgs sector, then
we expect $N \geq 3$ and two or more neutral scalars (with $CP=+1$)
would have degenerate masses $M_{h^o_j} \simeq 59 GeV$ (or: almost
degenerate, with
$\mid \triangle M_{h^o_j} \mid < \Gamma_{h^o_j}$).
Within such a model, it appears unlikely that the number of the
observed events can be explained without the mass degeneracy of scalars
$h^o_j$ (which don't couple to fermions).

The possibly ideal degeneracy of the $(N-1)$ (or: $N-l-k$) Higgs
scalars $h^o_j$ could result from an as yet unknown symmetry of the
Lagrangian.
\\
\\[1.cm]
{\bf Note Added}
After completing this paper, we were informed that the DELPHI Collaboration
{}~\cite{au} had recently observed (out of more than $10^6$ Z decays) two
such events with $M_{\gamma\gamma} \simeq 59 GeV$. Combining these results
with those of the L3 Collaboration leads again to the suggestion (within the
described scenario) that $(N-1) \geq 2$.
\\
\\[1.cm]
{\bf Acknowledgements.}
We thank E.~A.~Paschos and J.~Abraham for useful discussions
and comments on the subject. This work has been supported in part by the
the Deutsche Forschungsgemeinschaft (G.~C.~, Grant No.~PA254/7-1) and
the German Bundesministerium f\" ur Forschung und Technologie (M.~N.~, Grant
No.~BMFT 055DO9188; Y.~-L.~W.~, Grant No.~BMFT 055DO91P).


\newpage
\begin{thebibliography}{99}

\bibitem{L3}
The L3 Collaboration, Phys.~Lett.~B295 (1992) 337

\bibitem{Pohl}
M.~Pohl, talk given in Dortmund, 1 Dec.~1992

\bibitem{g}
J.~F.~Gunion, H.~H.~Haber, G.~L.~Kane and S.~Dawson, ``The Higgs Hunter's
Guide'' (Addison-Wesley, Reading, MA, 1990), p. 56

\bibitem{kw}
G.~Keller and D.~Wyler, Nucl.~Phys.~B274 (1986) 410

\bibitem{lhc}
Large Hadron Collider Workshop, Aachen, 1990, editors G.~Jarlskog, D.~Rein,
CERN 90-10, ECFA 90-133, Vol.~II, p.~428 (Z.~Kunszt, W.~J.~Stirling)

\bibitem{b}
V.~Barger, N.~G.~Deshpande, J.~L.~Hewett and T.~G.~Rizzo,
Argonne National Lab. preprint ANL-HEP-PR-92-102, Nov.~1992

\bibitem{au}
U.~Amaldi, private communication (DELPHI Note: ``A Study of Z Decays
into Two Leptons and Two Photons Using the DELPHI Detector'',
DELPHI Collaboration, CERN, 18 Dec.~1992)

\end{thebibliography}
\end{document}